\documentclass[12pt]{article}
\usepackage{amsmath}

\usepackage{amssymb}
\usepackage{amsmath,axodraw}
\usepackage{euscript}

\usepackage{epsfig}  
  
\usepackage{cite}  
  
\usepackage{alltt}  
\def \be{\begin{equation}}
\def \ee{\end{equation}}
\def \bea{\begin{eqnarray}}

\def \to {\rightarrow}
\def \eea{\end{eqnarray}}

\begin{document}
\begin{titlepage}
%Finding the CP Nature of the Higgs Boson at the NLC.
\begin{flushright}  
{\bf   
IFT-P.0xx/2003 \\
February 2003}  
\end{flushright}  

\hspace{1 mm}  
\vspace{1 cm}  
\begin{center}{\bf\Large {Higgs Production and Decay in the Little Higgs Model}}
\end{center}  
%\begin{center}{\bf\Large  Tevatron and its Three-Photon Signature.} \end{center}   
\vspace{0.3 cm}  
\begin{center}  
{\large\bf C. Dib$^{a,}$ \footnote{cdib@fis.utfsm.cl} ,~ R. Rosenfeld$^{b,}$ 
\footnote{rosenfel@ift.unesp.br} ~ and ~  A. Zerwekh$^{a,}$ \footnote{alfonso.zerwekh@fis.utfsm.cl}}
% A. Zerwekh$^{a,}$ \footnote{alfonso.zerwekh@fis.utfsm.cl} ,~ C.
%Dib$^{a,}$ \footnote{cdib@fis.utfsm.cl}  ~ and ~  R. Rosenfeld$^{b,}$
%\footnote{rosenfel@ift.unesp.br}}  
\vspace{0.3 cm}  
\\  
{\em $^a$ Department of Physics, Universidad T\'ecnica Federico Santa Mar\'{\i}a \\
Valpara\'{\i}so, Chile}\\  
{\em $^b$ Instituto de F\'{\i}sica Te\'orica - UNESP \\
 Rua Pamplona, 145, 01405-900 S\~ao Paulo, SP, Brasil}\\    
\vspace{0.5 cm}  
  
\end{center}

\centerline {\bf Abstract}
\vskip 1.0cm

We analyse the consequences of the little Higgs model for double Higgs boson
production at the LHC and for the partial decay width 
$\Gamma (H\to \gamma \gamma)$.   
In particular, we study the sensitivity of these processes in terms of the
parameters of the model. We find
that the little Higgs model contributions are proportional to $\left(\frac{v}{f}
\right)^4$ and hence do not change significantly either single 
or  double Higgs production at hadron colliders or 
$\Gamma (H\to \gamma \gamma)$ as compared to the standard model predictions.
%%%%%%%%%%%%%%%%%%
However, when interference and mixing effects are properly taken into account
these contributions increase to be of the order of   $\left(\frac{v}{f}
\right)^2$.
%%%%%%%%%%%%%%%%%%%%%%%%%%%
%We find that this process is not very sensitive to new
%physics as described by this model.
\bigskip

\vfill
\end{titlepage}

\newpage

\section{Introduction}

The presence of quadratic divergences in the loop processes for the scalar Higgs
boson self-energy in the standard model is responsible for the so-called hierarchy or 
fine-tuning problem. There is no natural way of protecting a light Higgs boson
from getting GUT scale contributions. This problem is solved in supersymmetric
extensions of the standard model, where the quadratic divergences are cancelled
by supersymmetric partners of the existing particles \cite{susy}. The hierarchy problem is
also absent in models in which the electroweak symmetry is dynamically broken,
since the scalar particles are not fundamental in these models \cite{techni}. 

Recently a new model was proposed which can solve in a natural way the hierarchy
problem of scalar Higgs boson in the standard model.  In this class of models,
called little Higgs models \cite{little}, 
the Higgs boson is a pseudo-Goldstone boson and its mass is protected by a 
global symmetry. The cancellations arise due to contributions of new particles
with the same spin.

The phenomenology of these models has been discussed with respect to indirect
effects on precision measurements and direct production of the new particles
\cite{pheno}.
In this letter we study yet another phenomenological consequence, the
contribution of new states to Higgs boson production and decay.

The little Higgs Lagrangian is given by the lowest order term of a non-linear
sigma model based on a coset $SU(5)/SO(5)$ symmetry:
\be
{ \cal L}_{\Sigma} = \frac{1}{2} \frac{f^2}{4} 
	{\rm Tr} | {\cal D}_{\mu} \Sigma |^2,
\label{Sigma}
\end{equation}
where the subgroup $[SU(2)\times U(1)]^2$ of $SU(5)$ is promoted to a local gauge
symmetry. The covariant derivative is defined as
         \begin{equation}
           {\cal D}_\mu \Sigma=  \partial_\mu\Sigma - i \sum_{j=1}^2\left( 
g_j( W_j\Sigma +  \Sigma W_j^T) + g'_j (B_j\Sigma + \Sigma B_j^T) \right).
	\end{equation}
To linearize the theory, one can expand $\Sigma$  in powers of ${1}/{f}$
around its vacuum expectation value $\Sigma_0$  
\begin{equation}
\Sigma = \Sigma_0 + \frac{2 i}{f} \left( \begin{array}{ccccc}
\phi^{\dagger} & \frac{h^{\dagger}}{\sqrt{2}} & {\mathbf{0}}_{2\times
2} \\
\frac{h^{*}}{\sqrt{2}} & 0 & \frac{h}{\sqrt{2}} \\
{\mathbf{0}}_{2\times 2} & \frac{h^{T}}{\sqrt{2}} & \phi
\end{array} \right) + {\cal O}\left(\frac{1}{f^2}\right),
\end{equation}
where $h$ is a doublet and $\phi$ is a triplet under the unbroken $SU(2)$.
The non-zero vacuum expectation value of the field 
$\langle \Sigma \rangle =\Sigma_0 $
leads to the breaking of the global $SU(5)$ symmetry to $SO(5)$ and also 
breaks the
local gauge symmetry $[SU(2)\times U(1)]^2$ into its diagonal subgroup, which is
identified with the standard model $SU_L(2)\times U_Y(1)$ symmetry group.
Following the notation of Han {\it et al.} \cite{pheno}, we will denote the
usual standard model gauge bosons mass eigenstates as 
$W^{\pm}_L$, $Z_L$ and $A_L$, where the 
subscript
$L$ denotes light in order to distinguish from the heavy states with mass 
of order $f$, denoted by
$W^{\pm}_H$, $Z_H$ and $A_H$.

The standard model fermions acquire their masses via the usual Yukawa
interactions. However, in order to cancel the top quark quadratic contribution
to the Higgs self-energy, a new-vector like color triplet fermion pair,
$\tilde t$ and $\tilde t^{\prime c}$, with quantum numbers
$({\mathbf{3,1}})_{Y_i}$ and $({\mathbf{\bar{3},1}})_{-Y_i}$
must be introduced. 
Since they are vector-like, they are allowed to have a
bare mass term which is {\it chosen} such as to cancel the quadratic 
divergence above scale $f$.

The coupling of the standard model top quark
to the pseudo-Goldstone bosons and the heavy colored fermions
in the littlest Higgs model is chosen to be 
\begin{equation}
{\cal{L}}_Y = {1\over 2}\lambda_1 f \epsilon_{ijk} \epsilon_{xy} \chi_i
\Sigma_{jx} \Sigma_{ky} u^{\prime c}_3 
+ \lambda_2 f \tilde{t} \tilde{t}^{\prime c}
+ {\rm h.c.},  
\label{yuk}
\end{equation}
where $\chi_i=(b_3, t_3, \tilde{t})$ and $\epsilon_{ijk}$ and $\epsilon_{xy}$
are antisymmetric tensors.
The new model-parameters
$\lambda_1,\ \lambda_2$ are supposed to be of the order of unity.

\setcounter{footnote}{0}

In terms of the mass eigenstates $\tilde{t}^{c}$ and $u^{ c}_3$, the term in the 
Lagrangian (\ref{yuk}) which describes the coupling of the new fermion to the
standard model (gauge eigenstate
\footnote{The standard model mass eigenstate Higgs will be denoted by $H$. The corrections due to
the difference between gauge and mass eigenstates are small (of the order $v^2/f^2$) and will be neglected in this
work. Likewise, we will neglect the mixing between $\tilde t$ and the top quark.})
 Higgs ($h^0$)is given by:

\begin{equation}
{\cal L}_{h-\tilde{t}} =
+ \frac{\lambda_1^2}{\sqrt{\lambda_1^2 + \lambda_2^2}}
\frac{1}{f} \left[ - \tilde t(h^+h^- + h^0h^{0*} +
 2\phi^{++}\phi^{--} + 2\phi^+\phi^- + 2\phi^0\phi^{0*}) \tilde t^c \right] .
\end{equation}
Notice that only a quartic coupling $ \tilde t h^0h^{0*}\tilde t^c$ is generated.

Another vertex that will be relevant to our analysis is $H W_L W_L$ and $H W_H
W_H$. It is an interesting characteristic of the model that they have opposite
signs. It must be so in order to cancel quadratic divergences in the Higgs self-energy.
 Neglecting mixing terms of higher order in $v/f$ one has \cite{pheno}:
\begin{eqnarray}
H W^+_{L \mu} W^-_{L \nu} & \Longrightarrow & \frac{i}{2} g^2 v g_{\mu \nu}
\nonumber \\
H W^+_{H \mu} W^-_{H \nu} & \Longrightarrow & -\frac{i}{2} g^2 v g_{\mu \nu}
\end{eqnarray}

We begin by investigating  the changes in the partial width 
$\Gamma (H \to \gamma \gamma)$ arising in this model. The partial width can be
written as \cite{collider}:
\begin{equation}
\Gamma (H \to \gamma \gamma) = \frac{G_F M_H^3}{8 \sqrt{2} \pi }
\left(\frac{\alpha}{\pi}\right)^2 |I|^2 ,
\end{equation}
where $|I|$ receives contributions from charged particles of spin $0$,$1/2$ and
$1$.
In the little Higgs model, there is an additional contribution in the loop from
the heavy vector boson $W^{\pm}_H$, which comes with the opposite sign of the
usual $W^{\pm}_L$. One could think that this would result in a partial 
cancellation between these two
contributions. However, since $M_{W_H} \simeq \frac{f}{v} M_{W_L}$, 
the contribution of the $W^{\pm}_H$ is suppressed by
a factor of roughly $\left(\frac{v}{f}
\right)^4$.

Notice that the new heavy fermion  does not contribute to this process and
charged scalar contributions are naturally small, since it must arise from the
Coleman-Weinberg effective potential in the scalar sector.

We now turn to Higgs boson production at the LHC in the little Higgs model.
The contribution to single Higgs production via gluon-gluon fusion is unchanged 
since a Yukawa coupling of 
the type $\tilde{t} \tilde{t^c} H$ does not exist in the linearized Lagrangian. 
However,
there is a contribution to Higgs pair production due to the quartic
$ \tilde t \tilde t^c H H$  term. We examine the possibility of 
observing this new 
contribution in Higgs boson pair production at hadron accelerators.

Gluon-gluon fusion is the dominant mechanism of standard model Higgs boson pair
production at the LHC \cite{dhiggs}. There is a top quark triangle and a top
quark box contributions. The differential partonic cross section in
the standard model can be written as, in the heavy quark limit:
\begin{equation}
\frac{d \hat{\sigma}(gg \to HH)}{d \hat{t}} = 
\frac{ G_F^2 \alpha_s^2}{256 (2 \pi)^3 } \left[2 \frac{M_H^2}{\hat{s}
- M_H^2} - \frac{2}{3} \right]^2,
\end{equation}
where $\hat{t}$ is the momentum transfer between an initial state gluon and a
final state Higgs boson. The total cross section is obtained by convoluting with
the gluon distribution function:
\begin{equation}
\sigma (pp \to HH) =    \int dx_1 dx_2 \; g(x_1,Q^2) g(x_2,Q^2) \hat{\sigma} (gg
\to HH) \theta (x_1 x_2 s - 4 M_H^2),
\end{equation}
where we have used the  Cteq6l1 leading order gluon distribution function
\cite{Cteq} with momentum scale $Q^2 = \hat{s}$. 
For the LHC, with $\sqrt{s} = 14$ TeV we obtain $\sigma (pp \to HH) = 38$ fb for
$M_H = 120$ GeV. With an expected luminosity of $10^{34}$ cm$^{-2}$ s$^{-1}$
\cite{lumLHC}  one would have of the order of $4000$ events in one year.

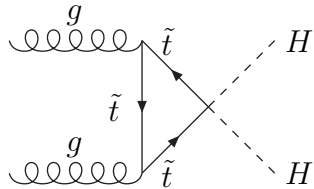
\begin{figure}[t]
  \begin{center}
    \begin{picture}(400,100)(0,0)
      \Text(210,75)[]{$H$}
      \Text(210,25)[]{$H$}
      \DashLine(200,75)(175,50){3}
      \DashLine(200,25)(175,50){3}
      \ArrowLine(175,50)(150,75)
      \Text(160,75)[]{$\tilde t$}
      \ArrowLine(150,25)(175,50)
      \ArrowLine(150,75)(150,25)
      \Text(160,25)[]{$\tilde t$}
      \Text(140,50)[]{$\tilde t$}
      \Gluon(150,75)(100,75){4}{5}
      \Text(125,85)[]{$g$}
      \Gluon(150,25)(100,25){4}{5}
      \Text(125,35)[]{$g$}
    \end{picture}
    \caption{Little Higgs Model contribution to the Higgs boson pair production at LHC.  }
    \label{2higgs}
  \end{center}
\end{figure}

In little Higgs models there is an extra contribution to this process shown in
figure (\ref{2higgs}). The amplitude for this process is given by:
\begin{equation}
{\cal M} (g^a g^b  \to HH) = 
g_{HH\tilde{t}\tilde{t}} \frac{\alpha_s}{\pi} \frac{\hat{s}}{6 m_{\tilde t}} \delta^{ab}
(\varepsilon_1 \cdot   \varepsilon_2),
\end{equation}
where $ \varepsilon_{1,2}$ are the gluon polarization vectors and the relevant
coupling constant is written as:
\begin{equation}
g_{HH\tilde{t}\tilde{t}}= - \frac{\lambda_1^2}{\sqrt{\lambda_1^2 + \lambda_2^2} } \frac{1}{f}
\approx \frac{1}{\sqrt{2} f}.
\end{equation}

This leads to the parton level cross section contribution from the little Higgs
model:
\begin{equation}
\hat{\sigma}_{LH} (gg \to HH) =
\frac{ g_{HH\tilde{t}\tilde{t}}^2 \alpha_s^2 \hat{s}}{9216 \pi^3 m_{\tilde t}^2} \sqrt{1-\frac{4
M_H^2}{\hat s}} \propto \frac{\hat s}{f^4}.
\end{equation}

The total cross section can be obtained by convoluting $\hat{\sigma}_{LH}$ with
the gluon distribution function. For $M_H = 120$ and $f = 2$ TeV, we obtain at 
the LHC the result
$\sigma_{LH} = 6 \times 10^{-3} $ fb, 4 orders of magnitude smaller than the
standard model. It only depends weakly on the Higgs mass and scales as $f^{-4}$. 
Since values of $f< 3.5$ TeV are excluded from precision measurements
\cite{pheno}, we conclude that the contribution of the little Higgs to the pair
production of the Higgs bosons seems to be unobservable at the LHC.

In conclusion, we have examined new contributions of the little Higgs model
to production and decay of the Higgs boson. We have found that the corrections
due to the new physics are at least of the order of $\left(\frac{v}{f}
\right)^4$. For $f> 3.5$ TeV, we expect small deviations of the order of
$10^{-3}$\%, probably too small to be detected at future accelerators.
\vspace{0.5cm}

{\it Note added: } 
After this paper was completed, another paper on a similar subject appeared
\cite{han}. The authors of this paper correctly included in the analysis 
interference and mixing effects which when properly taken into account results
in an increased contribution of the order of   $\left(\frac{v}{f}
\right)^2$.

\section*{Acknowledgments}

We would like to thank Tao Han for pointing out the contributions from
interference and mixing effects.
A.Z. and C.D. received partial support
from Fondecyt (Chile) grants No.~3020002 and 8000017,
respectively. R.R. would like to thank CNPq for partial financial support.
This work was also supported by Fundaci\'on Andes, Chile,
grant C-13680/4.                 

% Journal and other miscellaneous abbreviations for references
\def \arnps#1#2#3{Ann.\ Rev.\ Nucl.\ Part.\ Sci.\ {\bf#1} (#3) #2}
\def \art{and references therein}
\def \cmts#1#2#3{Comments on Nucl.\ Part.\ Phys.\ {\bf#1} (#3) #2}
\def \cn{Collaboration}
\def \cp89{{\it CP Violation,} edited by C. Jarlskog (World Scientific,
Singapore, 1989)}
\def \econf#1#2#3{Electronic Conference Proceedings {\bf#1}, #2 (#3)}
\def \efi{Enrico Fermi Institute Report No.\ }
\def \epjc#1#2#3{Eur.\ Phys.\ J. C {\bf#1} (#3) #2}
\def \f79{{\it Proceedings of the 1979 International Symposium on Lepton and
Photon Interactions at High Energies,} Fermilab, August 23-29, 1979, ed. by
T. B. W. Kirk and H. D. I. Abarbanel (Fermi National Accelerator Laboratory,
Batavia, IL, 1979}
\def \hb87{{\it Proceeding of the 1987 International Symposium on Lepton and
Photon Interactions at High Energies,} Hamburg, 1987, ed. by W. Bartel
and R. R\"uckl (Nucl.\ Phys.\ B, Proc.\ Suppl., vol.\ 3) (North-Holland,
Amsterdam, 1988)}
\def \ib{{\it ibid.}~}
\def \ibj#1#2#3{~{\bf#1} (#3) #2}
\def \ichep72{{\it Proceedings of the XVI International Conference on High
Energy Physics}, Chicago and Batavia, Illinois, Sept. 6 -- 13, 1972,
edited by J. D. Jackson, A. Roberts, and R. Donaldson (Fermilab, Batavia,
IL, 1972)}
\def \ijmpa#1#2#3{Int.\ J.\ Mod.\ Phys.\ A {\bf#1} (#3) #2}
\def \ite{{\it et al.}}
\def \jhep#1#2#3{JHEP {\bf#1} (#3) #2}
\def \jpb#1#2#3{J.\ Phys.\ B {\bf#1} (#3) #2}
\def \jpg#1#2#3{J.\ Phys.\ G {\bf#1} (#3) #2}
\def \mpla#1#2#3{Mod.\ Phys.\ Lett.\ A {\bf#1} (#3) #2}
\def \nat#1#2#3{Nature {\bf#1} (#3) #2}
\def \nc#1#2#3{Nuovo Cim.\ {\bf#1} (#3) #2}
\def \nima#1#2#3{Nucl.\ Instr.\ Meth. A {\bf#1} (#3) #2}
\def \npb#1#2#3{Nucl.\ Phys.\ B {\bf#1} (#3) #2}
\def \npps#1#2#3{Nucl.\ Phys.\ Proc.\ Suppl.\ {\bf#1} (#3) #2}
\def \npbps#1#2#3{Nucl.\ Phys.\ B Proc.\ Suppl.\ {\bf#1} (#3) #2}
\def \PDG{Particle Data Group, D. E. Groom \ite, \epjc{15}{1}{2000}}
\def \pisma#1#2#3#4{Pis'ma Zh.\ Eksp.\ Teor.\ Fiz.\ {\bf#1} (#3) #2 [JETP
Lett.\ {\bf#1} (#3) #4]}
\def \pl#1#2#3{Phys.\ Lett.\ {\bf#1} (#3) #2}
\def \pla#1#2#3{Phys.\ Lett.\ A {\bf#1} (#3) #2}
\def \plb#1#2#3{Phys.\ Lett.\ B {\bf#1} (#3) #2}
\def \pr#1#2#3{Phys.\ Rev.\ {\bf#1} (#3) #2}
\def \prc#1#2#3{Phys.\ Rev.\ C {\bf#1} (#3) #2}
\def \prd#1#2#3{Phys.\ Rev.\ D {\bf#1} (#3) #2}
\def \prl#1#2#3{Phys.\ Rev.\ Lett.\ {\bf#1} (#3) #2}
\def \prp#1#2#3{Phys.\ Rep.\ {\bf#1} (#3) #2}
\def \ptp#1#2#3{Prog.\ Theor.\ Phys.\ {\bf#1} (#3) #2}
\def \ppn#1#2#3{Prog.\ Part.\ Nucl.\ Phys.\ {\bf#1} (#3) #2}
\def \rmp#1#2#3{Rev.\ Mod.\ Phys.\ {\bf#1} (#3) #2}
\def \rp#1{~~~~~\ldots\ldots{\rm rp~}{#1}~~~~~}
\def \si90{25th International Conference on High Energy Physics, Singapore,
Aug. 2-8, 1990}
\def \zpc#1#2#3{Zeit.\ Phys.\ C {\bf#1} (#3) #2}
\def \zpd#1#2#3{Zeit.\ Phys.\ D {\bf#1} (#3) #2}


\begin{thebibliography}{99}


\bibitem{susy} See, {\it e.g.}, S.~P.~Martin in
{\sl ``Perspectives in Supersymmetry"}, edited by G.~L.~Kane, World
Scientific {\tt [arXiv:hep-ph/9709356]}.      

\bibitem{techni} For a recent review, see C.~T.~Hill and E.~H.~Simmons,
{\tt arXiv:hep-ph/0203079} and references therein (submitted to Phys. Rep.). 


\bibitem{little} 
N.~Arkani-Hamed, A.~G.~Cohen and H.~Georgi,
%``Electroweak symmetry breaking from dimensional deconstruction,''
Phys.\ Lett.\ B {\bf 513}, 232 (2001);
N.~Arkani-Hamed, A.~G.~Cohen, T.~Gregoire and J.~G.~Wacker,
%``Phenomenology of electroweak symmetry breaking from theory space,''
JHEP {\bf 0208}, 020 (2002);
N.~Arkani-Hamed, A.~G.~Cohen, E.~Katz, A.~E.~Nelson, T.~Gregoire and 
J.~G.~Wacker, 
JHEP {\bf 0208}, 021 (2002);
N. Arkani-Hamed, A.G. Cohen, E. Katz, and A.E. Nelson,
{\tt arXiv:hep-ph/0206021};
for a recent review, see {\it e.~g.}, M.~Schmaltz,
%``Physics beyond the standard model (Theory): Introducing the little  Higgs,''
{\tt arXiv:hep-ph/0210415} and J.~G.~Wacker, {\tt arXiv:hep-ph/0208235}. 


\bibitem{pheno}
C.~Csaki, J.~Hubisz, G.~D.~Kribs and P.~Meade, {\tt arXiv:hep-ph/0211124};
J.~L.~Hewett, F.~J.~Petriello and T.~G.~Rizzo,
{\tt arXiv:hep-ph/0211218};
G.~Burdman, M.~ Perelstein and A.~Pierce,
{\tt arXiv:hep-ph/0212228};
T.~Han, H.~E.~Logan, B.~McElrath and L.~-T.~Wang,  
{\tt arXiv:hep-ph/0301040}.

\bibitem{collider} See, {\it e.g.}, {\sl Collider Physics}, V.~D.~Barger and
R.~J.~N.~Phillips, Addison-Wesley (1987).

\bibitem{dhiggs} See, {\it e.g.}, T.~Plehn, M.~Spira and P.~M.~Zerwas,
Nucl. Phys. {\bf B479}, 46 (1996) (erratum ibid. {\bf 531}, 655 (98));
A.~Djouadi, W.~Killian, M.~Muhlleitner and P.~M.~Zerwas,
Eur. Phys. J. {\bf C10}, 45 (1999).

\bibitem{Cteq} Cteq Collaboration,  JHEP 0207 (2002) 012.

\bibitem{lumLHC}  A.~Clark, A.~Blondel and F.~Mazzucato,
ATL-PHYS-2002-029;
U.~Baur, T.~Plehn and D.~Rainwater,
Phys.\ Rev.\ Lett.\  {\bf 89}, 151801 (2002);
U.~Baur, T.~Plehn and D.~Rainwater,
{\tt arXiv:hep-ph/0211224}.

\bibitem{han} T.~Han, H.~E.~Logan, B.~McElrath and L.~-T.~Wang,  
{\tt arXiv:hep-ph/0302188}.








\end{thebibliography}
\end{document}